\documentclass[12pt]{article} 
\usepackage{epsfig}

\newcommand{\chandler}{Chandler-1998a,Chandler-1998b,Chandler-1998c,Chandler-1998d,Chandler-1998e}

\newcommand{\justdims}{Woolf-1998e,Zuckerman-1999,Zuckerman-2001}

\newcommand{\dphi}{\Delta \varphi}

\newcommand{\sdpathcite}{Elber-1990,Elber-1991,Karplus-1992,Theodorou-1993,Elber-1997}

\newcommand{\pacc}{p_{\mathrm{acc}}}

\newcommand{\targetcite}{Wollmer-1993,Diaz-1997,Schulten-1997}

\newcommand{\xbf}{{\mathbf{x}}}

\begin{document}
\title{Rapid Determination of Multiple Reaction Pathways in Molecular Systems:\\
The Soft-Ratcheting Algorithm}
\author{Daniel M. Zuckerman\footnote{Current address: Center for Computational Biology \& Bioinformatics, University of Pittsburgh, Pittsburgh, PA 15213, dzuckerman@ceoh.pitt.edu} $^\dagger$ and Thomas B. Woolf$^{\dagger \ddag}$\\
$^\dagger$Department of Physiology and  $^\ddag$Department of Biophysics,\\
Johns Hopkins University School of Medicine, Baltimore, MD 21205\\
}
\date{\today}
\maketitle

\begin{abstract}
We discuss the ``soft-ratcheting'' algorithm which generates targeted stochastic trajectories in molecular systems \emph{with scores corresponding to their probabilities}.
The procedure, which requires no initial pathway guess, is capable of rapidly determining multiple pathways between known states.
Monotonic progress toward the target state is \emph{not} required.
The soft-ratcheting algorithm is applied to an all-atom model of alanine dipeptide, whose unbiased trajectories are assumed to follow overdamped Langevin dynamics.
All possible pathways on the two-dimensional dihedral surface are determined.
The associated probability scores, though not optimally distributed at present, may provide a mechanism for estimating reaction rates.
\end{abstract}
\pagebreak
\section{Introduction}
Reaction paths in large molecular systems, such as biomolecules, provide critical information regarding structural intermediates (transitions states) and barrier heights.
The search for these paths has a long history in the applied math research commnity (e.g., \cite{Muller-1979}), as well as in the field of biomolecular computation \cite{\sdpathcite,Harvey-1993,Wollmer-1993,Diaz-1997,Schulten-1997,Pratt-1986,Chandler-1998a,Chandler-1998b,Chandler-1998c,Chandler-1998d,Chandler-1998e,Eastman-2000}.
Many approaches must start from an initial guess for the reaction path (such as a straight line between two states), effectively limiting the search to a single pathway.
On the other hand, ``targeted'' and ``steered'' MD approaches \cite{\targetcite} are capable of finding multiple pathways by repeated simulation (from differing initial conditions) forced to reach the desired end state.

The recently-introduced soft-ratcheting approach \cite{Zuckerman-2000} is also capable of ``blindly'' determining multiple reaction pathways.
It differs from the targeted and steered approaches in the following ways:
(i) monotonic progress toward the target state is not enforced, permitting a wider range of reaction pathways;
(ii) soft-ratcheting is applied in the context of stochastic dynamics, although this does not prevent the inclusion of explicit solvent molecules; and
(iii) a probability weight (``score'') is associated with each trajectory generated, which in principle also permits estmates of the reaction rates within the dynamic importance sampling (DIMS) formulation discussed by Woolf \cite{Woolf-1998e} and by Zuckerman and Woolf \cite{Zuckerman-1999,Zuckerman-2001}.
We note that reaction-rate estimates have not yet been produced by the soft-ratcheting algorithm, because such estimates require trajectories sampled with a near-optimal distribution (i.e., as would occur in unbiased dynamics; see below).

The soft-ratcheting procedure is simple and is motivated by the Metropolis algorithm \cite{Metropolis-1953} and the ``exponential transformation'' used in nuclear importance sampling (e.g., \cite{Booth-1979}).
Related methods include the ``weighted-ensemble Brownian dynamics" approach of Huber and Kim \cite{Huber-1996} and the ``CONTRA MD'' scheme of Harvey and Gabb \cite{Harvey-1993}.
The process is:
(a) generate an unbiased step;
(b) if the step moves toward the target state, accept it;
(c) if it moves away, accept it with a probability (i.e., ``softly'') that increases in the forward direction;
(d) repeat, while estimating the probability score for all accepted steps.
We emphasize that the non-monotonicity embodied in (b) and (c), and the existence of the score in (d) distinguish this method from previous multiple-pathway methods.
The guiding concept behind soft-ratcheting is \emph{not} to force the system (which necessarily perturbs the dynamics) but to try to allow the system to proceed in a possible, if unlikely, way.
Of course, rare stochastic events are just what we seek!

Note too that, unlike the trajectory sampling methods introduced by Pratt \cite{Pratt-1986} and pursued by Chandler and coworkers \cite{\chandler} as well as those of Eastman, Gr{\o}nbech-Jensen and Doniach \cite{Eastman-2000}, the \emph{overall} effect of the soft-ratcheting algorithm is non-Metropolis in nature (despite the motivation): 
\emph{trajectories do not evolve from one another and are statistically independent}.
The Metropolis idea is only used to ensure that a given trajectory successfully reaches the target state.
In this important sense, the soft-ratcheting algorithm comes under the independent-trajectory rubric of the DIMS method \cite{\justdims}.

\section{\label{sec:theory}Theory}
\subsection{Generating Paths}
In essence there is no more theoretical underpinning to producing soft-ratcheted trajectories than that already sketched in the Introduction:
using a physically-but-arbitrarily chosen \emph{acceptance probability function} for step increments, one accepts all forward steps, and backward steps are accepted with a probability which decreases the more ``negative'' the increment.
See Fig.\ \ref{fig:p-accept}.
Here, the forward direction is simply some vector in configuration space that points from the initial to the target state --- perhaps a dihedral angle in a dihedral transition.
The algorithm is sufficiently robust (see Results section) that advance knowledge of the reaction path and the true reaction coordinate is not necessary.

When generating a series of soft-ratcheted crossing events in a single long trajectory, it is convenient to use a simple \emph{threshold} device \cite{Zuckerman-2001}.
This means only that trajectories are permitted to perform unbiased dynamics in small regions near the ``centers'' of the beginning and end states, and biased (i.e., soft-ratcheted) dynamics begin only when the threshold is reached.
The idea is to allow the trajectory to explore different parts of the stable states, with an eye toward finding exit points to different pathways.
Such exploration, of course, must take place within the limits of available computer time!
As noted below, our use of the threshold requires further investigation and optimization, though it appears to perform the task of permitting exploration of alternative exit points from a stable state.

\subsection{Scoring Paths}
It is only when one wishes to associate a score with a trajectory that some analysis must be undertaken.
The dynamic importance sampling (DIMS) approach requires the probability score for use in rate estimates, moreover.
Specifically, the probability score used in DIMS calculations is the ratio of two quantities \cite{\justdims}:
(i) the probability that the given trajectory would occur in an unbiased simulation --- a known quantity; and
(ii) the probabilty that the given trajectory was produced in the biased (e.g., soft-ratcheting) simulation.
Further details of the full DIMS formulation may be found in Refs.\ \cite{\justdims} and are beyond the scope of the present report.
Here we focus solely on computing the probability that the soft-ratcheting algorithm produced a given trajectory (ii), which unfortunately does not follow directly from the simple acceptance probability used to generate the trajectory.

This section gives full details of generating the probability score (i.e., ratio) required by DIMS.
Briefly, however, assume progress towards the target is measured in terms of a scalar ``distance,'' $\varphi$, which is larger at the target state than the initial:
each step corresponds to an increment $\dphi$, with positive increments moving toward the target.
From any starting configuration $\xbf_{n-1}$, one can define the \emph{unbiased} distribution of $\dphi$ increments
$ p_{\dphi}(\dphi; \xbf_n | \xbf_{n-1}) $, which is simply the projection of the more fundamental distribution of configurations, $\xbf_n$, onto the $\dphi$ coordinate.
The distribution of $\dphi$ increments typically is nearly Gaussian with a mean which may be either positive or negative.
However, once certain backward steps are rejected due to the acceptance function in the soft-ratcheting procedure (specified below), the $\dphi$ distribution is shifted forward in a non-trivial way to become the biased distribution, 
$ b_{\dphi}(\dphi; \xbf_n | \xbf_{n-1}) $.
Estimating the ratio of values of these two distributions for every accepted step (though not the entire distributions) is the task at hand.

The multi-dimensional case reduces to a simple scalar description in terms of $\dphi$ increments, but we include it for completeness.
We assume (although it is not necessary for the formalism) that the initial and final states of interest in our molecule do not require the full all-atom configuration $\xbf$, but rather a subset of coordinates, say, $\{\phi_1, \phi_2, \ldots \}$.
If the target point is the ``center'' of state B, say, $\{ \phi^B_i \}$, then one can always measure the distance to that point,
\begin{equation}
\label{phi-distance}
d_B(\{ \phi_i \}) = \left [  \sum_i \left( \phi_i - \phi^B_i \right)^2  \right ]^{1/2} \;,
\end{equation}
where it may be necessary to consider the closest distance if the $\phi_i$ coordinates represent true angles.
For a step from $\xbf_{n-1}$ to $\xbf_n$, one can then define a one-dimensional change in distance by
\begin{equation}
\label{delta-phi}
\dphi(n\!-\!1 \rightarrow n) = d_B(\{ \phi_i^{(n)} \}) - d_B(\{ \phi_i^{(n-1)} \})
\; .
\end{equation}
In essence, since distance from the target is always a scalar quantity, one need only consider a one-dimensional description to estimate probability scores.

\begin{figure}[here]
\begin{center}
\epsfig{file=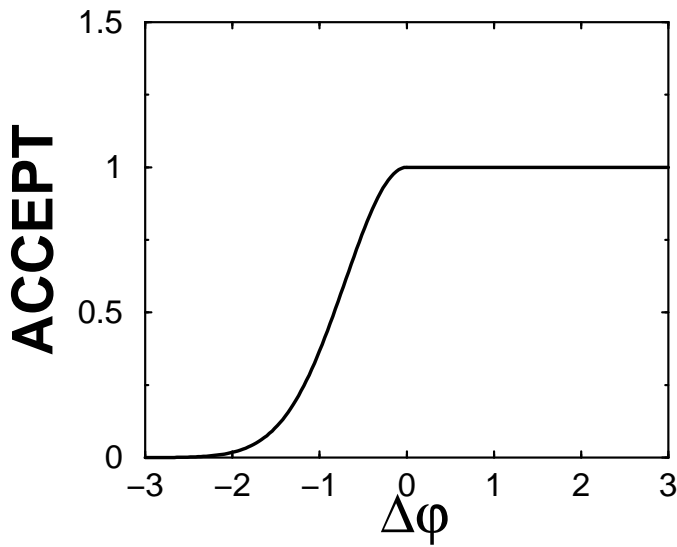, height=3in}
\end{center}
\vspace*{1cm}
\caption{\label{fig:p-accept}
An example acceptance function for use in the soft-ratcheting algorithm.
The acceptance probability $\pacc$ for a given step is plotted against the ``distance'' toward the target in angle space, $\dphi$.
Steps toward the target ($\dphi > 0$) are always accepted, while steps away from the target ($\dphi < 0$) are accepted with probability less than one.
Thus, trajectories are not forced toward the target, but ``softly'' ratcheted.
}
\end{figure}

The acceptance function for $\dphi$ increments is very simple and is specified by the simulator.
The function used in the present work is illustrated in Fig.\ \ref{fig:p-accept} and is written
\begin{equation}
\label{p-accept}
\pacc(\dphi) =
\left\{ 
  \begin{array}{ll}
  1 & \mbox{if } \dphi > 0 \\ 
  \exp{\left[ -|\dphi/\dphi_0|^2 \right]} & \mbox{if } \dphi < 0 \; ,
  \end{array}
\right.
\end{equation}
where $\dphi_0$ is a parameter which controls the width of the (backwards) decay depicted in Fig.\ \ref{fig:p-accept}.
The gradual decay to zero is the ``softness'' of soft-ratcheting:
many backwards steps will be accepted.

With $\pacc$ specified, the final task toward generating the required probability score is to consider the relation between the unbiased and soft-ratcheted (biased) distribution.
The probability (density) that the soft-ratcheting algorithm will generate a given $\varphi$ increment, $b_{\dphi}$, is proportional to the product of the unbiased probability of generating the increment, $p_{\dphi}$, and the acceptance probability, $\pacc$:
\begin{equation}
\label{single-step-prob}
b_{\dphi}(\dphi) = {\cal N}^{-1} p_{\dphi}(\dphi) \, \pacc(\dphi) \; ,
\end{equation}
where ${\cal N} < 1$ is the required normalization factor, given by the fraction of steps initiated at $\xbf_{n-1}$ which would be accepted by the soft-ratcheting procedure.
As noted, the biased distribution, $b_{\dphi}$, has been shifted forward in the $\varphi$ direction because the acceptance function $\pacc$ partially suppresses backward steps.

The desired probability score for a single step is then the ratio deriving from (\ref{single-step-prob}), namely,
\begin{equation}
\label{single-step-ratio}
\mbox{single-step ratio} = \frac{p_{\dphi}(\dphi)}{b_{\dphi}(\dphi)} 
	= \frac{{\cal N}}{\pacc(\dphi)} \; .
\end{equation}
To truly calculate the normalization factor $\cal N$, one would have to initiate a large number of steps from the point $\xbf_{n-1}$ and compute the fraction accepted by the soft-ratcheting acceptance function.
As that procedure would be very computationally expensive, we instead use the sequence of nearby \emph{attempted} steps, both accepted and rejected, to estimate the probability that soft-ratcheted steps were accepted in a given local region of configuration space.
The final score is simply the product of the single-step scores (\ref{single-step-ratio}).

\section{\label{sec:results}Results}
\begin{figure}[here]
\begin{center}
\epsfig{file=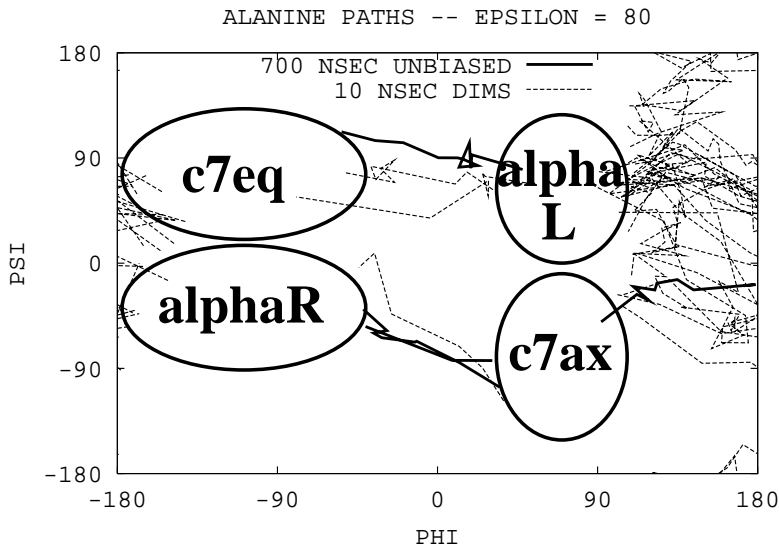, height=3in}
\end{center}
\vspace*{1cm}
\caption{\label{fig:short-time}
Rapid generation of crossing trajectories with the soft-ratcheting algorithm.
The figure shows both crossing trajectories generated by unbiased simulation (dark lines) and those generated by the soft-ratcheting algorithm in a fraction of the uniased simulation time (dashed lines, ``DIMS'').
The potential is AMBER94 \cite{amber} as encoded in the Molecular Modelling Tool Kit \cite{mmtk} for an all-atom representation of alanine dipeptide, and the unbiased trajectories were generated using overdamped Langevin dynamics.
}
\end{figure}

\begin{figure}[here]
\begin{center}
\epsfig{file=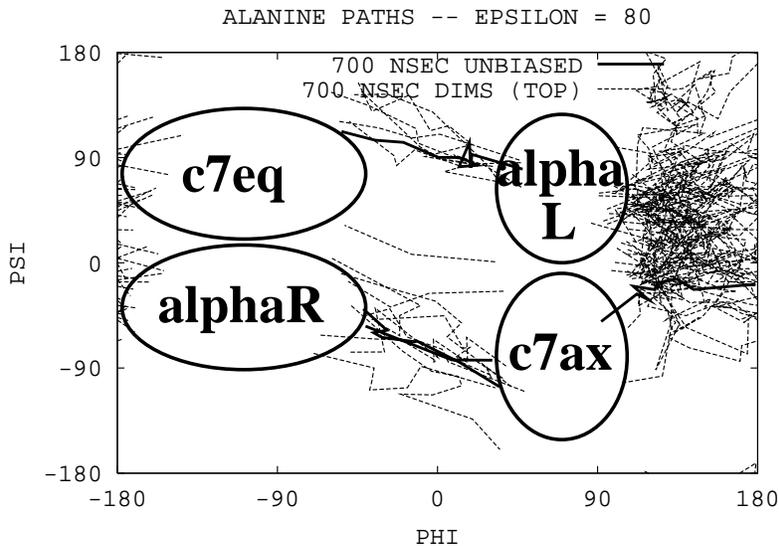, height=3in}
\end{center}
\vspace*{1cm}
\caption{\label{fig:long-time}
Accuracy of top-scoring crossing trajectories generated with the soft-ratcheting algorithm.
The figure shows both crossing trajectories generated by unbiased simulation (dark lines) and the \emph{top-scoring} trajectories  generated by the soft-ratcheting (dashed lines, ``DIMS'').
Note that the large ensemble of soft-ratcheted trajectories appear to better explore the full pathway ``valleys.'' 
Data are from simulations of equal length, with the same potential and dynamics as in Fig.\ \ref{fig:short-time}.
}
\end{figure}

The results of this preliminary report may be summarized in three points:
(i) the soft-ratcheting algorithm is capable of generating reaction pathways \emph{rapidly} --- in a fraction of the time which would be required by unbiased simulation: see Fig.\ \ref{fig:short-time};
(ii) the scores associated with each crossing trajectory permit the generation of a most-important \emph{ensemble} of events as in Fig.\ \ref{fig:long-time}, which can give more detailed information about the full ``valley'' of the pathway; and
(iii) the associated scores, in principle, permit \emph{rate estimates} within the dynamic importance sampling formulation \cite{\justdims}.

In Figure \ref{fig:short-time}, one sees the rapidity with which the soft-ratcheting algorithm generates crossing trajectories.
The same three pathways are found in 1/70th of the simulation time.  
In absolute terms, the 10 nsec.\ of simulation time used in generating the soft-ratcheting trajectories appears quite long;
however, this time may be significantly reduced by adjusting the threshold level (see Sec. \ref{sec:theory}) from the preliminary value used to generate the depicted results.

Figure \ref{fig:long-time} illustrates the capacity of the soft-ratcheting algorithm to generate an ``important'' (highly weighted) ensemble of crossing events.
The large set of trajectories shown in the figure clearly gives a better description of the pathway valleys than the sparse events generated by unbiased simulation.

Figure \ref{fig:long-time} also demonstrates the agreement between the weight estimate discussed previously (used to select the depicted trajectories) and the unbiased results.
The higher-weighted trajectories coincide strongly with the unbiased events.
The large cluster of soft-ratcheted trajectories in the region $110 < \phi < 180$ deserves comment.
Because there is only a single unbiased event in that region, it is not obvious whether the relatively widely dispersed soft-ratcheted trajectories are ``correct'' --- i.e., whether such an ensemble of trajectories would be found in a long unbiased simulation, with many events in the region.
Examination of the adiabatic energy surface (not shown) does indicate that the channel in question is indeed significantly wider than the two pathways crossing $\phi = 0$, though perhaps not quite to the extent suggested by the soft-ratcheting trajectories of Fig.\ \ref{fig:long-time}.

\section{Future Research}
Several means of improving the soft-ratcheting procedure are possible, of which we mention two.
First, to increase the speed with which transition trajectories are generated --- really, to decrease the waiting interval between crossing events --- one can reduce the size of the threshold region (Sec.\ \ref{sec:theory}) in which purely \emph{unbiased} dynamics are performed.
The threshold region was intended to permit trajectories to explore a multiplicity of potential ``exit points'' from the stable state.
However, the ``softness'' of the soft-ratcheting algorithm should, by itself, permit a substantial degree of this kind of exploration, and it may be possible to use a very small threshold region.

Second, a more optimal (i.e., higher-scoring) ensemble of trajectories presumably can be obtained by systematic estimation of parameter $\dphi_0$.
In fact, the promising preliminary results presented in Sec.\ \ref{sec:results} were based on an \emph{ad-hoc} choice.
It is a simple matter to study in more detail an unbiased distribution of $\dphi$ increments, and then use this data to systematically inform the choice of $\dphi_0$.
Moreover, one can imagine attempting to bias trajectories forward in a focussed conical region of dihedral angles \cite{Booth-1979}, rather than simply according to (hyper)planes of constant $\dphi$.

Ultimately, it will also be important to compare the soft-ratcheted paths (which presumably represent the stochastic dynamics in a faithul way) with those generated by explicitly-solvated molecular dyanmics simulation.
That is, how does the addition of explicit solvent alter the paths?
Of course, this comparison will only be possible in small molecules like the alanine dipeptide and other small peptides, but it will provide a crucial validation of the technique.

\section{Summary and Discussion}
We have given motivation and details for the ``soft-ratcheting'' algorithm \cite{Zuckerman-2000} for determining reaction pathways in molecular systems governed by stochastic dynamics.
The method generates independent transition trajectories which will not be trapped in a single channel (pathway), and hence is capable of finding multiple channels.
Although a final state is always targeted on average, the algorithm permits ``backward'' steps with a suppressed probability.
The trajectories are thus ratcheted forward, but only softly: see Fig.\ \ref{fig:p-accept}.
The capacities of the approach were demonstrated in Figs.\ \ref{fig:short-time} and \ref{fig:long-time} for an all-atom model of the alanine dipeptide molecule evolving according to overdamped Langevin dynamics with the AMBER potential \cite{amber}.

Beyond rapidly generating multiple pathways, as other existing approaches are presently able to do \cite{\targetcite}, the soft-ratcheting algorithm has the potential also to estimate reaction rates and free energy differences via the dynamic importance sampling (DIMS) framework \cite{\justdims}.
The soft-ratcheting algorithm associates a score (see Sec.\ \ref{sec:theory}) with each transition trajectory it generates.
The scores, in turn, may be used in principle to estimate kinetics and free energy differences.
At present, however, we note that initial results showed that further parameterization and/or refinement of the algorithm are necessary before efficiency can be obtained in rate and free energy calculations.

\section*{Acknowledgments}
We gratefully acknowledge funding provided by the NIH (under grant GM54782), the AHA (grant-in-aid), the Bard Foundation, and the Department of Physiology.
Jonathan Sachs offered many helpful comments on the manuscript.

\end{document}